\documentclass[11pt]{article}
\usepackage[T1]{fontenc}
\usepackage[latin1]{inputenc}
\usepackage{graphicx}

\usepackage{moriond,epsfig}
\usepackage{floatflt}

\def\be{\begin{equation}}
\def\ee{\end{equation}}
\def\bea{\begin{eqnarray}}
\def\eea{\end{eqnarray}}

\begin{document}
 \def\bone{B^{(1)}} \def\etal{{\textit{et al.~}}} \def\eg{{\textit{e.g.~}}} \def\ie{{\textit{i.e.~}}} \def\DM{dark matter~} \def\DE{dark energy~} \def\GC{Galactic centre~} \def\susy{SUSY~}

\hfill {\tt PCCF-RI-0505}
\vspace*{4cm}

\title{The Maximal Neutrino Flux\\
 from Neutralino Annihilation in the Galactic Center}

\author{J. Orloff}

\address{Laboratoire de Physique Corpusculaire,\\
Universit{\'e} Blaise Pascal,\\
24 av. des Landais, F-63177 Aubi{\`e}re Cedex }

\maketitle \abstracts{ We discuss a robust and fairly model-independent
  upper bound on the possible neutrino flux produced by neutralino
  annihilation in the center of our galaxy, and show that its detection
  with present or future neutrino telescopes is highly improbable. This
  bound is obtained by relating the neutrino flux to the gamma flux that
  would be produced in the same annihilation processes, for which
  measurements do exist.  }

\section{Introduction: Neutralino Dark Matter}

A large number of cosmological observations on scales ranging from
galactic or cluster sizes up to the cosmological horizon itself, clearly
show that known matter (baryons) and radiation (photons, and neutrinos
in a certain sense) gravitationally coupled by general relativity
fail to provide a complete description of the observed Universe. In
this scientifically challenging situation, some necessarily new ingredient
is needed. Modifications of gravity have been proposed, that can reasonably
cope with the galactic (newtonian) scales but require more work before
being extended to the largest (relativistic) ones.

Another perhaps less drastic and more testable possibility, is to keep
gravity intact and just imagine some new neutral (and thus dark)
matter\cite{Bertone:2004pz}. After all, the progresses of particle physics
in the last fourty years have provided countless examples of new particles
that could easily incarnate such new dark matter, except that their
lifetimes are extremely short on a cosmological timescale.  Another new
particle $\chi$ is thus needed. It should be the least ad-hoc possible, and
possess a cosmological lifetime. This requires an extremely small effective
coupling
$\alpha_{\chi}\doteq\Gamma_{\chi}/m_{\chi}<2
\times10^{-42}\mathrm{GeV}/m_{\chi}$.
Since $n-$loop processes generically give much too large contributions of
the order $\alpha_{QED,EW,QCD}^{n}$, a new symmetry is also needed to
guarantee their vanishing. For these reasons, supersymmetry emerges
together with R-parity to stabilize the lightest supersymmetric particle.
The neutralino, a mixture of the SUSY partners of scalars and electroweak
gauge bosons, is a well studied and fairly predictive dark matter
candidate. Its relic density $\Omega_{\chi}$ for instance, if determined
from CMB measurements, fixes the annihilation cross-section
$\sigma_{\chi\chi}^{ann}\sim\Omega_{\chi}^{-1}$ which in turn puts
constraints on the particle physics model. This is because neutralinos once
reached a status a thermodynamical equilibrium, which erased all memory of
initial conditions. For the plots below\cite{Bertone:2004ag}, we considered
CMSSM (a.k.a mSugra) models with the following parameters:
$50\mathrm{GeV}<m_{0}<4\mathrm{TeV}$,
$50\mathrm{GeV}<m_{1/2}<2\mathrm{TeV}$, $A_{0}=0$, $\tan\beta=5, 20, 35$.
We also considered the deviations from gaugino universality at the Grand
Unification scale $M_{2}|_{GUT}=0.6m_{1/2}$ or $M_{3}|_{GUT}=0.6m_{1/2}$
(instead of $1m_{1/2}$), which have the most important effects on
annihilation.

\section{Indirect Detection: Uncertainties and Interest of the
Galactic Centre}

\begin{floatingfigure}[r]{0.44\textwidth}%
\noindent\includegraphics[%
  bb=0bp 0bp 567bp 470bp,
  clip,
  width=0.44\textwidth]{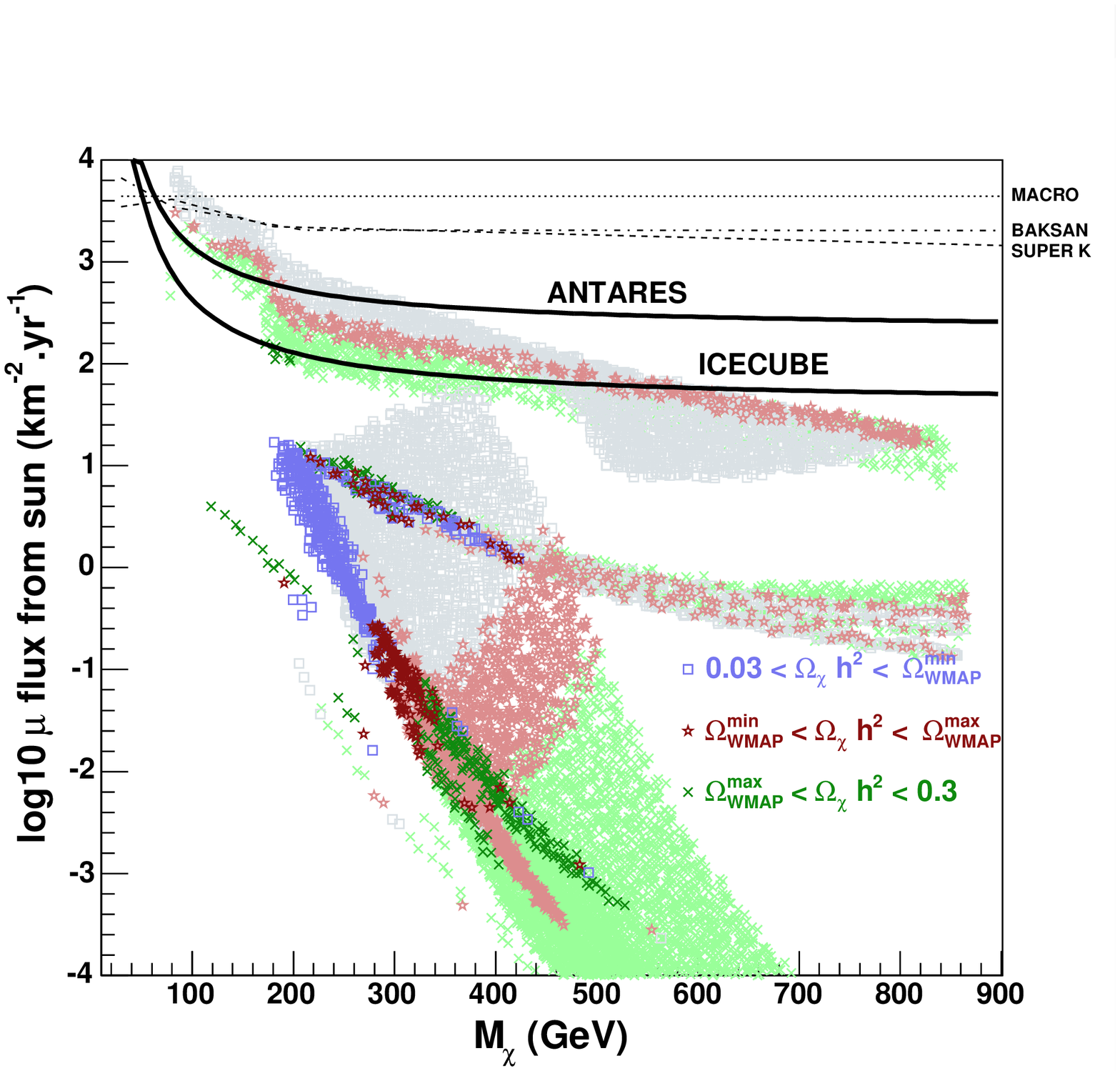}\\
\includegraphics[%
  bb=0bp 4bp 567bp 480bp,
  clip,
  width=0.44\textwidth]{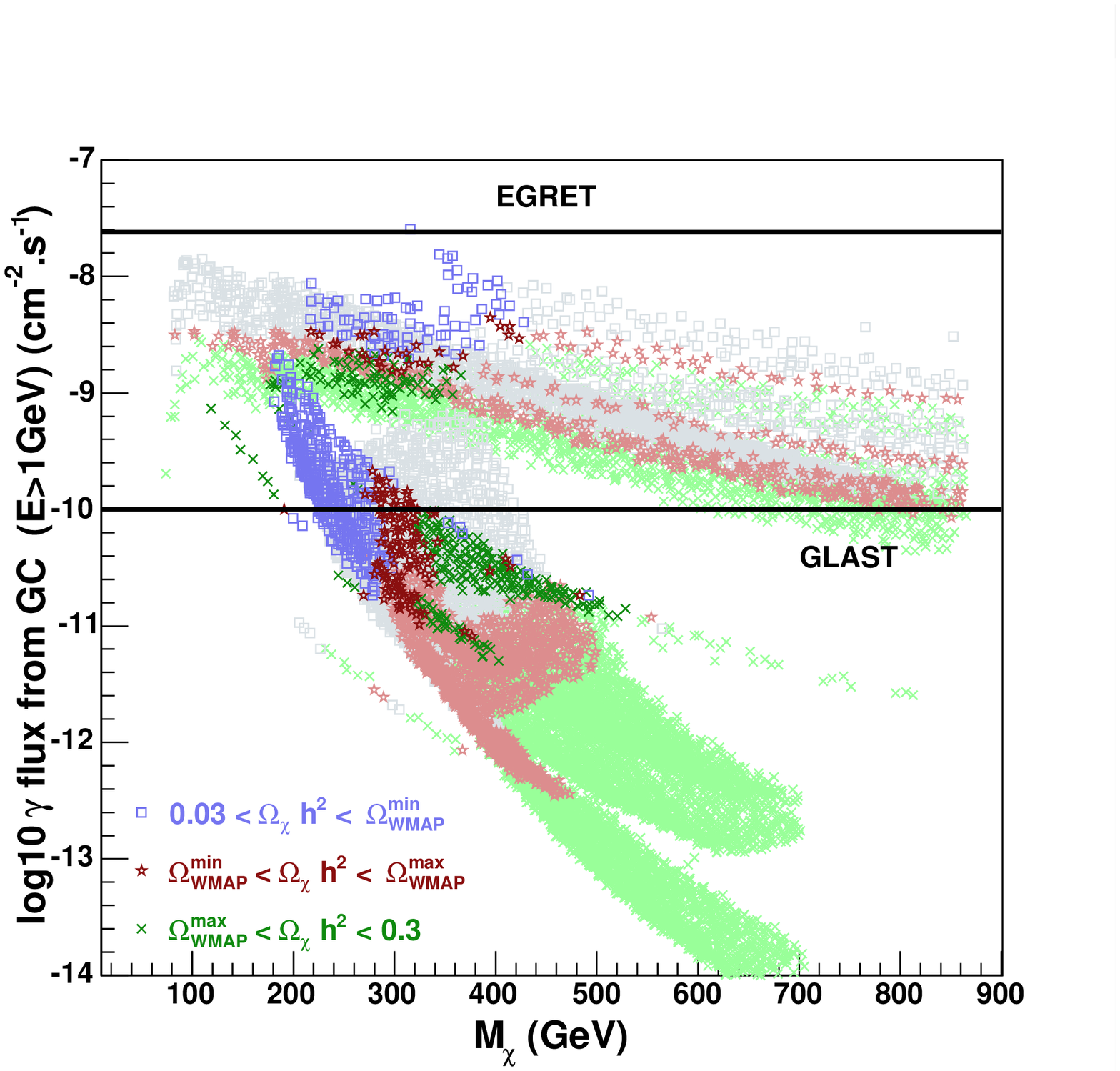}
\caption{Indirect detection signals and experimental sensitivities: (top)
neutrino-induced up-going muon flux above 25 GeV from neutralinos
annihilating in the Sun; (bottom) $\gamma$ flux above 1 GeV from
the GC, assuming a NFW profile with $J=1300$; here and
below, shades paler than in the legend denote models with a low, SM-like
anomalous dipole moment of the muon $\delta_{\mu}^{susy}<8.1\,10^{-10}$.}
\label{IDfreenorm}\end{floatingfigure}%
The indirect detection of dark matter is first a hunt for places where
dark matter can be sufficiently concentrated to start its self-annihilation
again. The annihilation products (gammas, neutrinos, antiparticles...)
can then be looked for. Naively, one would guess that fixing the annihilation
cross-section to comply with WMAP measurements, also fixes the indirect
detection signal. This not quite so for three reasons.

First, the annihilations that occur at freeze-out involve higher kinetic
energies than the later annihilations for indirect detection, which
occur essentially at rest. Certain annihilation processes are then
forbidden for symmetry reasons. The channel dominating the annihilation
rate, and thus the controlling parameters, can be different.

Second, depending on the annihilation product looked for, and on the
experimental sensitivity that can be achieved, different annihilation
channels may become relevant\cite{Bertin:2002ky}. For instance, in
neutrino indirect detection, high energy neutrinos are both easier
to detect and less numerous in the background. This makes the annihilation
channels which proceed via a pair of gauge bosons (each of which can
deposit half its energy in a neutrino) more relevant than channels
proceeding through a pair of light quarks (which mostly fragment into
hadronic cascades with little energy left for a neutrino).

Finally, the flux of annihilation products goes like the square of
the neutralino density. For neutralinos trapped in the gravitational
wells of celestial bodies like the earth or the sun, this is fixed
by how fast elastic collisions on the matter making these bodies can
slow down neutralinos below the escape velocity. Although the cross
section for elastic collision {\sl on matter} is related to the cross
section for annihilation {\sl into matter} by crossing symmetry,
the dominating amplitude and the relevant parameter may differ.

All these effects pile up to induce a large variability in indirect
detection signals. This is illustrated on top Fig.~\ref{IDfreenorm} for the
neutrino signal from annihilation of neutralinos captured inside the Sun:
constraining the relic density within the 13\% WMAP\cite{Spergel:2003cb}
uncertainties $\Omega_{M}h^{2}=0.135_{-0.009}^{+0.008}$ still leaves a 7
orders of magnitude room for the signal.

Our Galactic Center (GC) provides another, even deeper potential well than
the sun, and thus possibly larger indirect detection fluxes.  However, the
variability in these signals resulting from the unknown dark matter density
profile is even larger. Indeed this profile can only be inferred from the
dynamics of sources orbiting our Galactic Center, which feel the total mass
inside radius $R$: $M_{\chi}(R)=4\pi\int^{R}dr\, r^{2}\rho_{\chi}(r)$,
where the small $r$ contribution is strongly supressed. On the other hand,
indirect detection signals are proportional to the squared density
integrated along the line of sight in a direction $\psi$, often
parametrized by
$J(\psi)=(8.5\,\mathrm{kpc})^{-1}(0.3\mathrm{GeV/cm}^{3})^{-2}
\int_{l.o.s}ds\,\rho_{\chi}^{2}(r(s,\psi))$,
which in the direction of the GC can vary from 30 (isothermal profile) to
over $10^{5}$ (Moore profile) or even more in the presence of an accretion
spike on the GC Black-Hole. Even fixing this $J$ factor to an intermediate
NFW value of 1300 (as in bottom Fig.~\ref{IDfreenorm}), the first two reasons
above leave a three orders of magnitude range for the photon indirect
detection signal from the GC. Larger values of $J$ can bring certain
particle models within the reach of EGRET.  For the neutrino indirect
detection signal from the GC, the situation is qualitatively the same,
except that the larger gap between signal and experimental sensitivities
requires larger values of $J$ for observation. However such large values
would imply a huge photon signal, that has not been seen.
\\

\section{A Model Independent Upper Bound}
The observed photon flux from the GC clearly gives an upper bound on the
photon flux from neutralino annihilation in the GC, which can be translated
into an upper bound on the neutrino flux from the GC.\cite{Bertone:2004ag}
Indeed for a given \DM candidate and particle physics contents, the ratio
between the number of photons and the number of neutrinos emitted per
annihilation is known. We can thus estimate the neutrino flux from the GC
associated with a gamma-ray emission reproducing the EGRET data. Finally we
can convert the flux of neutrinos into a flux of muons, produced by
neutrinos interactions with the rock around detectors on Earth, in order to
compare with experimental sensitivities.

\begin{figure}
\includegraphics[%
  bb=0bp 5bp 555bp 470bp,
  clip,
  width=0.48\textwidth]{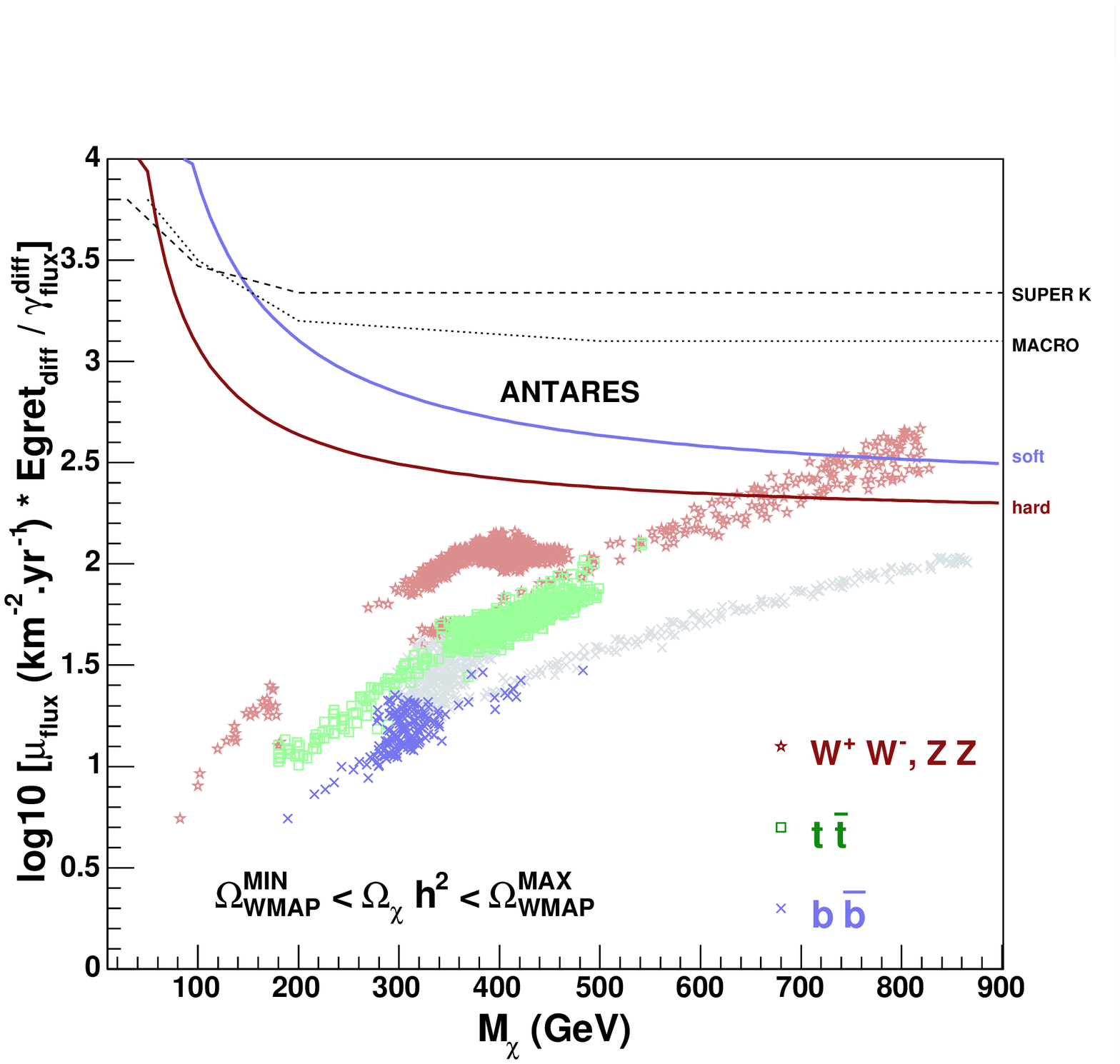}\hfill{}\includegraphics[%
  bb=0bp 5bp 555bp 470bp,
  clip,
  width=0.48\textwidth]{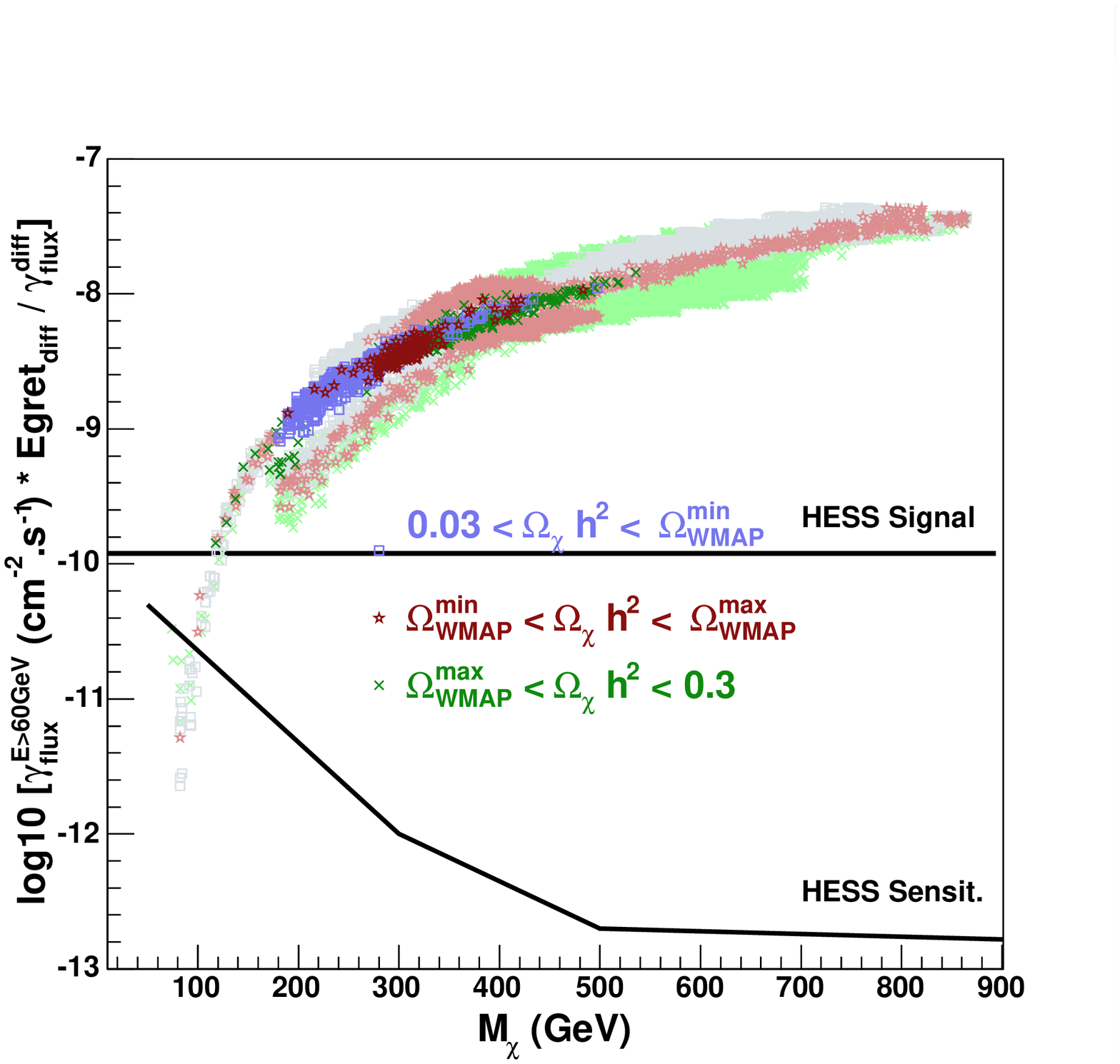}

\caption{(left) Neutrino-induced muon flux from the \GC normalized to EGRET,
sorted by leading ($\equiv BR>0.4$) annihilation channel for the
preferred WMAP relic density; (right) the photon flux (above 60 GeV)
normalized to EGRET, together with the planned HESS sensitivity above
60 GeV and the actual signal (with a higher threshold). }

\label{rescaledNuGamma}
\end{figure}

The rescaled flux of muons $\phi_{\mu}^{\textrm{{norm}}}(>E_{th})$ will
thus be given by 
\begin{equation}
  \phi_{\mu}^{\textrm{{upper}}}(>E_{th})=
  \frac{\phi_{\mu}^{\textrm{{NFW}}}(>E_{th})}{
    \phi_{\gamma}^{\textrm{{NFW}}}(E_{*})}
  \,\phi_{\gamma}^{\textrm{{EGRET}}}(E_{*})\label{eq:norm}\end{equation}
where the label NFW reminds that NFW profiles have been used to compute the
(profile-independent) flux ratio, and $E_{*}$ is the energy at which we
decide to normalize the flux to the gamma-ray data (in our case
$E_{*}=2$GeV). The results are shown in the left
  Fig..~\ref{rescaledNuGamma}.
The ratio in (\ref{eq:norm}) is by construction
independent of $J$, but it also turns out to be rather model-independent:
for a given neutralino mass, it spans less than a factor 10, which can be
traced to the dominant annihilation channel. The comparison with Antares sensitivity shows that only the
highest neutralino masses above 650 GeV can possibly be detected in the
Galactic centre.

However, for such large masses, a higher choice of photon energy $E_{*}$ would
allow to tighten this upper bound. To crudely evaluate how much, we
show in Fig.~\ref{rescaledNuGamma}-right the photon flux above 60~GeV
that should come together with the EGRET flux if it were 100\% due
to neutralino annihilation in the GC. The fact that HESS actually
sees a smaller flux implies that at most 1\% can be attributed to
neutralinos above 650 GeV, which lowers the possible neutrino flux
by as much for these masses. An update of Fig.~\ref{rescaledNuGamma}-left
including the most recent photon fluxes from the GC remains to be
done, but the resulting neutrino upper bound should flatten out for
neutralinos above 300 GeV, leaving little hope for neutrino telescopes.

\section{Discussion and loopholes}

If neutrinos are nevertheless observed above the given fluxes, their
interpretation as due to neutralino annihilation is problematic. The only
possibility would then be to invoke selective absorption of the photons by
electrons in the GC. However the photon mean free path being
$\lambda_{\gamma}\approx100\mathrm{kpc}
(E_{\gamma}/1\mathrm{GeV})(10^{5}\mathrm{cm}^{-3}/n_{e})$, this would
require huge electron densities.

Switching to other dark matter candidates, like Kaluza-Klein
resonances\cite{Servant:2002aq}, allows to increase the hard neutrino flux
above the bound presented here. Indeed, neutralinos cannot annihilate into
a hard neutrino anti-neutrino pair because of their Majorana nature.
However, no natural candidate annihilates \emph{only} into neutrinos, so
that the present bound can only be relaxed by a factor
$\approx1/(1-BR(\nu\bar{\nu}))$.

Finally, one may wonder if astrophysical sources other than dark matter
annihilation could provide detectable neutrino fluxes, within the realm of
the Standard Model. This question has recently been
adressed\cite{Crocker:2004nk} similarly using relations between the photon
and neutrino fluxes, with a more positive conclusion.

\section*{Acknowledgments}

We gladly aknowledge support from the French GdR 2305 {}``Supersym{\'e}trie'',
and thank the organisers for a most stimulating and informative conference.
In the short space available, we must also apologize to our coauthors
G.~Bertone, E.~Nezri, and J.~Silk for oversimplifying the arguments
and to other authors for the limited references.


\begin{thebibliography}{1}
\bibitem{Bertone:2004pz}For a recent review, see G.~Bertone, D.~Hooper and J.~Silk, Phys.\ Rept.\ \textbf{405}
(2005) 279.
\bibitem{Bertone:2004ag}G.~Bertone, E.~Nezri, J.~Orloff and J.~Silk, Phys.\ Rev.\ D
\textbf{70} (2004) 063503 .
\bibitem{Spergel:2003cb}D.~N.~Spergel \textit{et al.}, Astrophys.\ J.\ Suppl.\ \textbf{148}
(2003) 175.
\bibitem{Bertin:2002ky}V.~Bertin, E.~Nezri and J.~Orloff, Eur.\ Phys.\ J.\ C \textbf{26}
(2002) 111 \& JHEP \textbf{0302}, 046 (2003).
\bibitem{Servant:2002aq}G.~Servant and T.~M.~Tait, Nucl.\ Phys.\ B \textbf{650} (2003)
391.
\bibitem{Crocker:2004nk}R.~M.~Crocker, F.~Melia and R.~R.~Volkas, Astrophys.\ J.\textbf{\ 622}
(2005) L37 .
\end{thebibliography}
\end{document}